# Laser power transmission in space: Plasma-based power cell

*Li Lin and Michael Keidar**

School of Engineering and Applied Science, George Washington University. 800 22nd St. NW, Washington, DC 20052, USA

*Michael Keidar. keidar@gwu.edu

ABSTRACT

Laser power beaming offers a route to space energy delivery, but semiconductor laser photovoltaic receivers face thermalization, joule heat, and radiative recombination waste, etc. Here we propose a gas-phase plasma power cell that converts vacuum-ultraviolet photons into electrical output through xenon photoionization and magnetically biased charge separation. Particle-in-cell Monte Carlo simulations of a low-pressure xenon chamber driven by a 58.4 nm pulsed laser predict a steady-state laser-to-electrical conversion efficiency of 82.25% at 2000 W/m$^2$ average incident power. Energy accounting closes to 1%, with 7.52% photon escape, 8.95% boundary loss, and 1.29% chamber-stored energy. Parameter scans over bias voltage, magnetic field, and gas pressure identify photon absorption and electron confinement as controlling design factors. These results are a proof-of-concept gas-phase receiver architecture for space laser power beaming.

**TOC GRAPHICS**

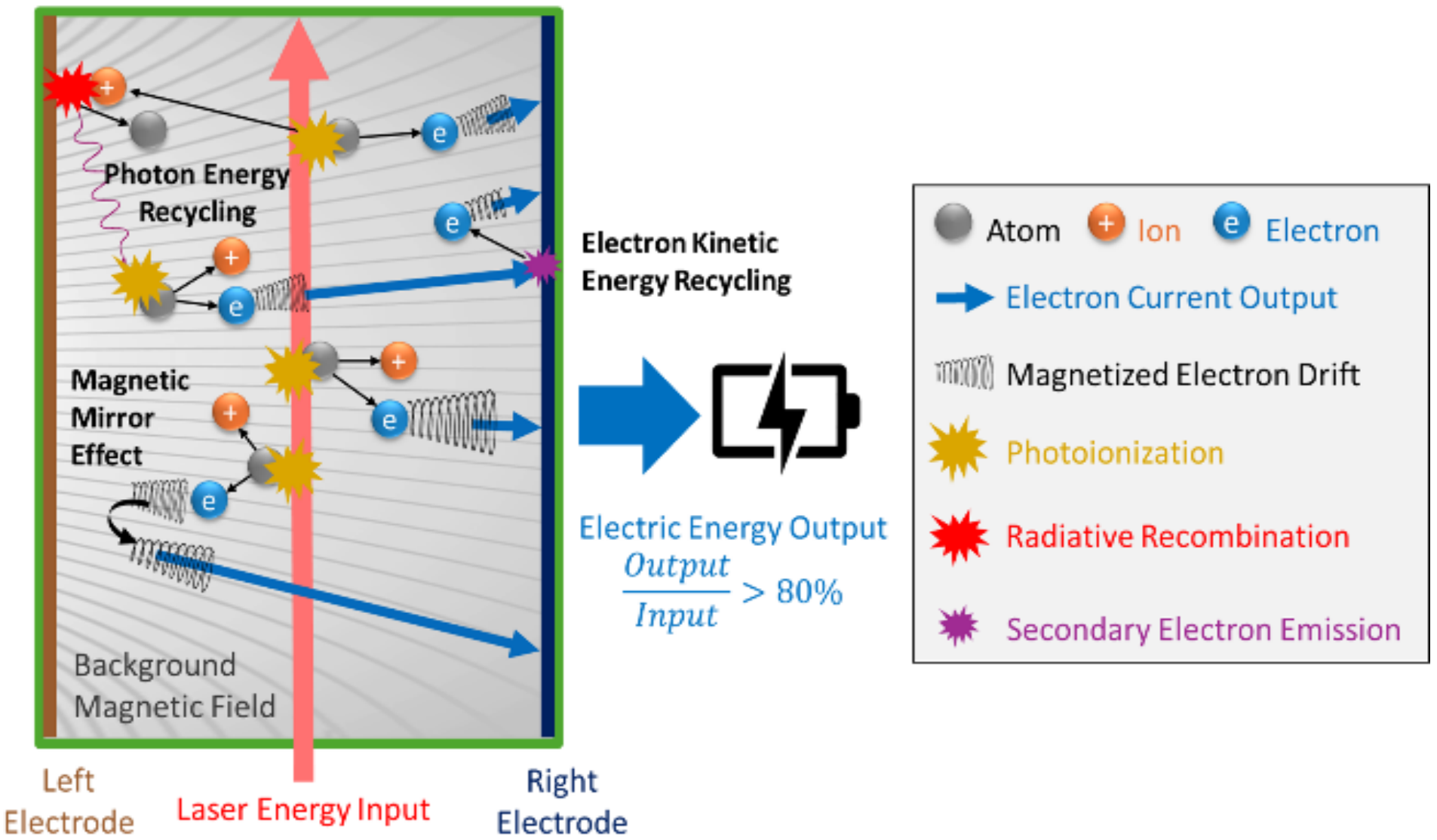

MAIN TEXT

Remote power delivery is currently a critical capability for the next-generation space systems, enabling sustained operation of high-power propulsion, distributed spacecraft networks, and deep-space missions where direct solar harvesting is limited. Solar panels remain the dominant power source for spacecraft, from CubeSats to space stations. In near-Earth orbit, solar irradiance is sufficient, however, practical performance is reduced by orbital geometry, orbital illumination, and Shockley-Queisser limit [1,2]. These constraints limit high-power propulsion, lunar operations, and deep-space exploration. Laser-based power beaming has therefore attracted increasing interest as an alternative approach for continuous and targeted energy delivery, due to a more accurate bandgap matching. However, current semiconductor laser photovoltaic receivers are fundamentally limited in efficiency by Joule heating, radiative recombination and also suffer from material degradation under high-intensity radiation environments, with the latest demonstrated efficiencies typically at about 65% [3–9].

In this work, we propose a different approach: a laser-driven plasma energy receiver, in which incident photons are absorbed through photoionization in a low-pressure gas rather than in solid-state junctions. Figure 1 (a) illustrates the operating principle of the plasma-based laser receiver considered in this numerical proof-of-concept. Incident laser photons are absorbed in a low-pressure gas chamber through photoionization, producing electron–ion pairs directly in the plasma volume. The generated photoelectrons carry excess kinetic energy above the ionization threshold of the laser photons. In contrast to a solid-state photovoltaic absorber, where excess carrier energy is rapidly transferred to the lattice and is dissipated, in plasma, this energy initially remains in the free-electron population. An externally applied magnetic field then shapes the electron transport. Although the magnetic field (a permanent magnet) does not supply energy, it changes electron trajectories. In particular, the magnetic-mirror geometry provides a net probability that energetically accessible electrons can be collected by the output electrode on the right, while reflects part of the electron population moving toward the virtually grounded electrode on the left. The electrons reach the right electrode thus deliver electrical energy to the external circuit against the charging potential (1$^{st}$ order Thevenin model) on the right electrode. The external circuit connects the right electrode of plasma chamber to the cathode of spacecraft battery, and the virtually grounded left electrode is connected to the anode of the spacecraft battery. Two diodes are added in the external circuit to ensure that the battery is being charged rather than discharging energy to the plasma.

The major advantage of plasma-based photovoltaic system is that the plasma can partially recycle the energy usually are lost in semiconductor-based photovoltaic devices. First, secondary electron emission (SEE) at the electrodes converts a fraction of electron-impact energy into newly emitted electrons, reducing the effective wall-loss channel. Second, radiative recombination provides another recovery route: recombination photons can be reabsorbed by the gas through excitation or ionization, and the resulting excited states can transfer part of their energy back to electrons through subsequent radiative or collisional processes [10,11]. This is also known as the radiative trapping effect [12]. These processes do not imply complete recovery of walls or radiative losses, but this is a unique energy recycling probability that are not significant in semiconductors. The photon escape, boundary losses, and residual chamber energy are explicitly tracked in the energy budget that will be discussed in later paragraphs. The overall receiver operation is therefore governed by the coupling among photoionization, magnetized electron confinement, electrode collection, and partial energy recycling, which are resolved and investigated in this work using a self-consistent particle-in-cell with Monte-Carlo collisions (PIC-MCC) simulation.

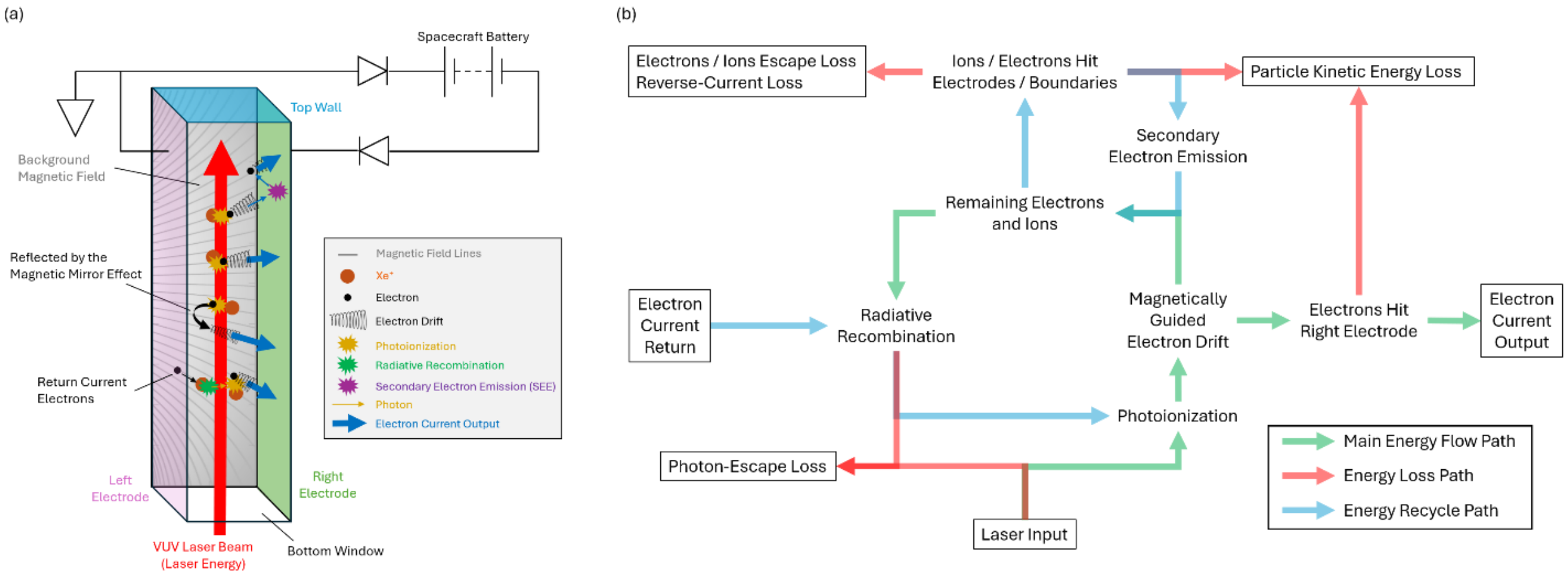

**Figure 1.** Concept and operating principle of the plasma-based laser photovoltaic system. (a) The schematics showing magnetically guided electron drift and charge separation in the plasma chamber for current output. (b) The energy pathway of plasma-based laser photovoltaic system.

The same processes can be viewed as an energy-flow network, as summarized in Figure 1(b). The incident laser energy first enters the chamber through photon absorption. The absorbed portion is converted into charged-particle kinetic energy, electrostatic potential energy, and ionization-energy reservoirs, whereas unabsorbed photons are recorded as photon-escape loss. The useful output channel is the collection of electrons at the biased output electrode, where the particle energy is converted into electrical work in the external circuit. Competing channels include particle escape, residual kinetic-energy deposition at boundaries, reverse ion current, and energy remaining in the plasma chamber. The recovery pathways shown in Figure 1(b) do not represent additional external energy sources; rather, they redirect part of the internally stored or otherwise lost energy back into mobile charged particles or photons that can further participate in photoionization, excitation, and charge extraction. This energy-flow picture provides the physical basis for the quantitative energy accounting used in the simulation.

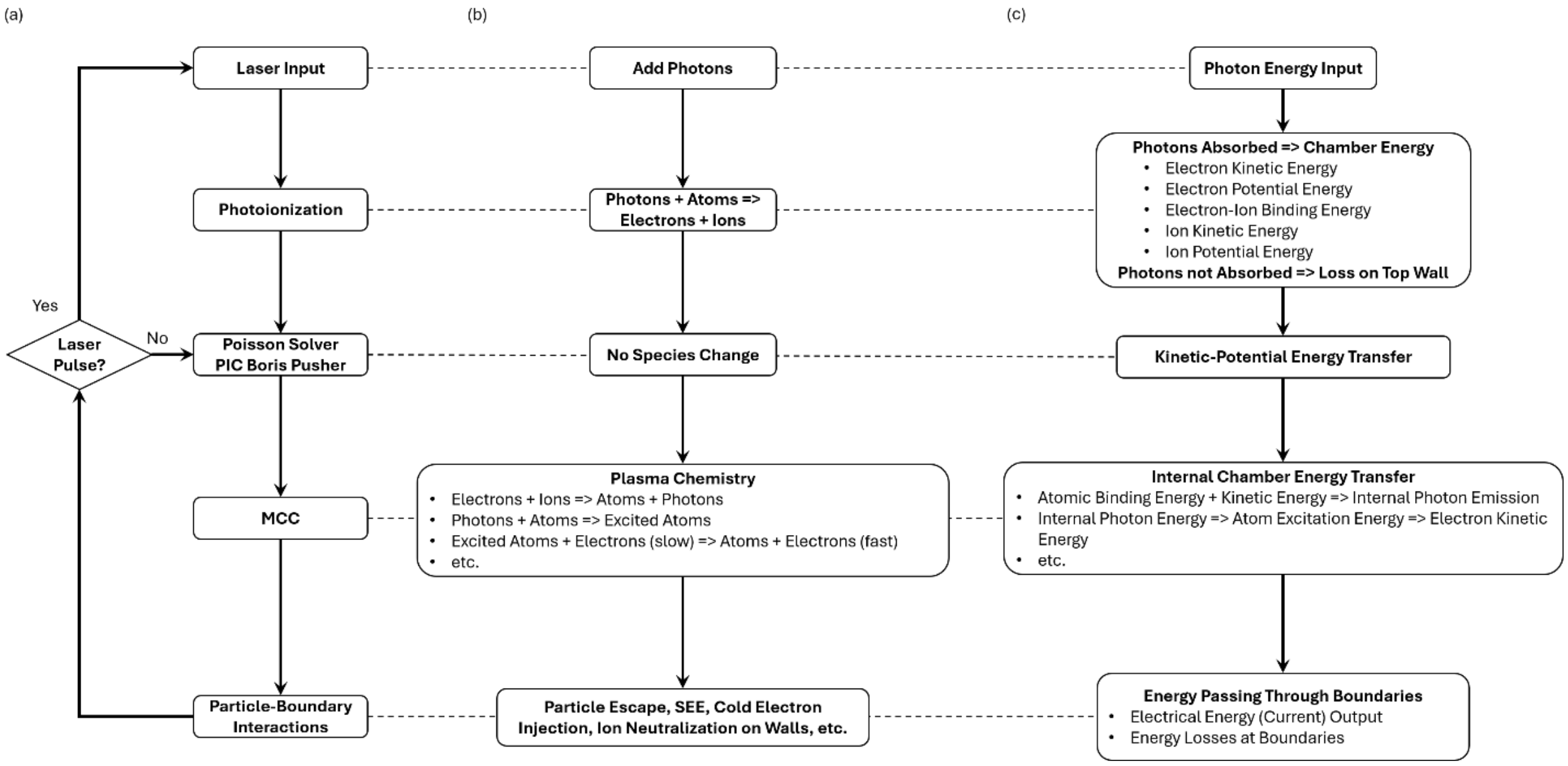

**Figure 2.** Detailed workflow linking (a) PIC-MCC simulation modules, (b) species bookkeeping, and (c) energy bookkeeping.

Figure 2 summarizes the PIC-MCC simulation flow chart and how the particles and energies are processed in each simulation module. First, photon absorption creates electron-ion pairs and transfers the absorbed photon energy into electron kinetic energy and atomic ionization energy.

During the field-solve and Boris-pusher step, species identities are unchanged and particle energy is redistributed between kinetic and electrostatic potential energy. MCC chemical events then transfer energy among charged particles, atoms, excited states, and internally emitted photons. At the boundary update, escaping photons and particles are recorded as losses, collected electrons contribute to electrical output, and return-current electrons are reinjected as cold electrons without external kinetic-energy input. This bookkeeping ensures that electrical output, photon escape, boundary loss, and chamber-stored energy are mutually exclusive categories in the control-volume energy budget.

The overall energy conversion efficiency is thus defined as the accumulated energy output divided by the accumulated laser energy input as shown in Eq. (1):

$$\eta = \frac{E_{Output}}{E_{Laser}} \tag{1}$$

where $E_{Output}$ is the energy output to the external circuit, $E_{Laser}$ is the laser input energy. The output energy can be quantified from the extracted electron current. The energy output is the result of magnetized electrons drift against the charging voltage potential barrier and reached the right electrode as

$$E_{Output} = UI_{net}\mathrm{d}t \tag{2}$$

where $U$ is the charging voltage, $I_{net}$ is the net current output, and $\mathrm{d}t$ is the time step. The loss at the boundaries contains the loss of particles at the bottom window, and the kinetic energy losses at the walls, defined as Eq. (3):

$$E_{BL} = E_{Left} + E_{Right} + E_{Top} + E_{Bottom} \tag{3}$$

where $E_{Left}$ is the energy loss at the left electrode, $E_{Right}$ is the energy loss at the right electrode, $E_{Top}$ is the energy loss at the top wall, $E_{Bottom}$ is the energy loss at the bottom aperture. These terms are specifically defined as:

$$E_{Bottom} = \sum_j \left(\frac{1}{2} m_j |\vec{v}|_j^2 + \varphi_j q_j\right) + E_{AB_Bottom} \tag{4a}$$

$$E_{Top} = \frac{1}{2} \sum_i \left(m_i v_{y,i}^2\right) \tag{4b}$$

$$E_{Right} = \sum_k \left(\frac{1}{2} m_k |\vec{v}|_k^2 - E_{SEE,k}\right) \tag{4c}$$

$$E_{Left} = \sum_l \left(\frac{1}{2} m_l |\vec{v}|_l^2 - E_{SEE,l}\right) \tag{4d}$$

where the indices $i$, $j$, $k$, and $l$ represent the particles that reach the top wall, exit the bottom aperture, reach the right electrode, and reach the left electrode, $E_{AB_Bottom}$ is the atomic binding energy loss due to the particle escape through the bottom aperture, $v_y$ is the vertical velocity of a particle, and $E_{SEE}$ is the energy used to generate SEE. To incorporate the SEE contribution at the electrode boundaries, we use the Vaughan-type empirical SEY model[13,14] (see Supporting Information).

The detailed setup of this work is the following. First, the laser wavelength is 58.4 nm (He $^1P^o$ => $^1S$) with an average power density is 2000 W/m$^2$ on average and 2 MW/m$^2$ per pulse. The laser pulse frequency is 1 GHz with the pulse width at 1 ps. The chamber pressure is $2.5 \times 10^{-4}$ atm filled with pure xenon, and is a 2D simulation domain with the size at 1 mm × 15 cm with a background magnetic field maximum value is 0.5 T at $x = 0$ mm and $y = 7.5$ cm. The left wall is a grounded electrode for the return electron current, and the right wall is an electrode at -30 V which is the charging voltage of a spacecraft cell. Such a high aspect-ratio geometry provides a sufficiently long optical path for efficient photon absorption. The bottom of the chamber is a molecular-flow vacuum ultra-violet (VUV) aperture in this work, because the laser at such wavelength can hardly pass through lens. However, the opening does not mean a significant leak of xenon and other particles because its size is much less than the mean-free-path (for the details of gas leak, see Supporting Information). The top boundary is treated as an idealized charge-retaining, non-absorbing wall. When a particle reaches this boundary, its wall-normal velocity component is set to zero, while its tangential velocity, charge, and species identity are retained. The removed normal kinetic energy as shown in Eq. (4b). Electrostatics at the top boundary is described by a homogeneous Neumann condition of zero normal electric field. Charge accumulation is represented explicitly by the retained particles in the near-wall PIC cells rather than through a separate surface-charge variable.

Please note that although 58.4 nm photoionization produces primary electrons with approximately 9.1 eV excess kinetic energy, collection does not require every electron to traverse the full 30 V bias. An electron generated at $\boldsymbol{r}_0$ encounters the local barrier $e[\varphi(\boldsymbol{r}_0) - U]$, electrons born near the output electrode can therefore be collected directly, whereas others may contribute after magnetized transport and internal energy redistribution. The 30 V bias is thus the terminal voltage rather than a universal single-particle barrier.

Figure 3 shows the main results of the simulation. Figure 3(a) presents the cumulative energy balance, in which the output, boundary-loss, chamber energy storage, and photon-escape terms are time-integrated energy fluxes. At 75 ns, the system reached a steady state meaning that the accumulated laser input energy, output energy, losses, chamber energy and the photon escapes as shown in Figure 3(a) have all become linear with a constant slope, although they all have small oscillation due to the laser pulses. Overall efficiency has also converged to a constant at 82.25% as defined by Eq. (1), with energy loss at all the boundaries at about 8.95% of the total energy, and chamber stored 1.29% of the total energy. There are 7.52% of input laser photons are not absorbed and wasted at the top wall. The slope in Figure 3(a) gives the overall temporal averaged power output at steady state of about 1645 W/m$^2$ (electrical output power density normalized by the laser-aperture area) while the laser input power is 2000 W/m$^2$ in the chamber.

Figure 3(b) gives a more detailed energy analysis showing, at steady state in average over all absorbed laser photons, where the energy from a single absorbed laser photon is distributed. The input laser photon is 58.4 nm meaning 21.26 eV as shown on the top of the figure. In which, 19.14 eV is output, 0.4 eV becomes the electron kinetic energy in the chamber, 0.058 eV becomes the electron potential energy in the chamber, $1.83\times10^{-4}$ eV becomes the ion kinetic energy, -0.45 eV becomes the ion potential energy (the right electrode is -30V, meaning the potential in the chamber is negative), 0.28 eV is the atomic binding remaining in the chamber, 0.75 eV is lost on the right electrode, 1.227 eV is lost on the left electrode, almost 0 eV is lost on the top wall (at steady state, the top wall surface charge prevent more electrons to reach there), and 0.078 eV is lost through the bottom aperture. Please note that Figure 3(b) is showing the energy distribution of an absorbed photon, it leads to a 19.14 eV / 21.26 eV ~ 90% efficiency, while the overall efficiency, as shown in Figure 3(a), of the system also contains the loss of non-absorbed photons.

Figure 3(c) shows the current density output at the steady state. Because the laser power density and the output current density are normalized by different surfaces, their numerical values are related through the chamber aspect ratio. The incident laser power density is normalized by the bottom laser-aperture area, $A_{laser}$, whereas the output current density is normalized by the side-electrode area, $A_{\mathrm{electrode}}$. For the present $1\ \mathrm{mm} \times 15\ \mathrm{cm}$ chamber, $A_{\mathrm{electrode}}/A_{\mathrm{laser}} = 150$. Consequently, an electrode current density of approximately $0.366\ \mathrm{A\ m^{-2}}$ corresponds to an incident-area-normalized electrical output of approximately $1645\ \mathrm{W\ m^{-2}}$at $U = 30$ V.

To verify the energy balance in the entire system and simulation, we quantified the error:

$$E_{error} = E_{Laser} - E_{Chamber} - E_{BL} - E_{PE} - E_{Output} \quad (5)$$

where $E_{Laser}$ is the accumulated laser energy input, $E_{Chamber}$ is the energy stored in the chamber, $E_{PE}$ is the energy loss due to photon escape. The chamber energy, as defined in Eq. (6):

$$E_{Chamber} = \sum_{\xi} \left(\frac{1}{2} m_{\xi} |\vec{v}|_{\xi}^{2} + \varphi_{L,\xi} q_{\xi}\right) + \frac{1}{2} \sum_{\xi} \left(\varphi_{P,\xi} q_{\xi}\right) + E_{AB} \quad (6)$$

where $\xi$ is the subscript of particles in the chamber, $m$ is the particle mass, $\vec{v}$ is the particle velocity, $\varphi_L$ is the charge-free Laplacian potential solved based on the electrode boundary condition, $\varphi_P$ is the self-consistent space-charge potential with $\varphi_P = \varphi - \varphi_L$ and $\varphi$ is the Poisson equation solution, $q$ is the charge of the particle (equaling to 0 for atoms), $E_{AB}$ is the total atomic binding energy in the chamber. At steady state, the resulting energy balance error of the entire system is shown in Figure 3(d) where the error w.r.t. the laser input energy is at about 1%. Such a small error means that the overall energy balance is well obeyed in the simulation, with no missing losses and double-counted energies. The energy accounting is based on mutually exclusive energy categories, such that each energy carrier contributes to only one term in Eq. (5), preventing double counting during the entire simulation. In the simulation, whenever an energy-conversion event occurs (photoionization, recombination, wall interaction, or particle extraction), the original energy carrier is removed from the corresponding category before the newly generated energy carrier is added to its new category. An example of radiative recombination: an electron and ion will be deleted, including their kinetic energies and potential energies in the simulation program, while a xenon atom and a photon are generated with the photon having the energy of ionization threshold. In the Supporting Information, the error analysis on the external electrical output, stored chamber energy, photon escape, and boundary losses closes the energy budget, confirming the self-consistency of the simulation and with no double-counting in the budget.

Figure 4 identifies a coupled operating window for the plasma receiver. Increasing the bias raises the extracted energy per collected electron until the accompanying reduction in collection offsets the voltage gain. A stronger magnetic field improves collection by suppressing electron loss to the grounded electrode, without supplying energy. Increasing xenon pressure primarily shortens the photon-absorption length and raises the absorbed fraction, which bounds the overall efficiency. These mechanism-consistent trends show that the 82.25% result is not an isolated operating point but arises from coordinated optical, magnetic, and electrical design.

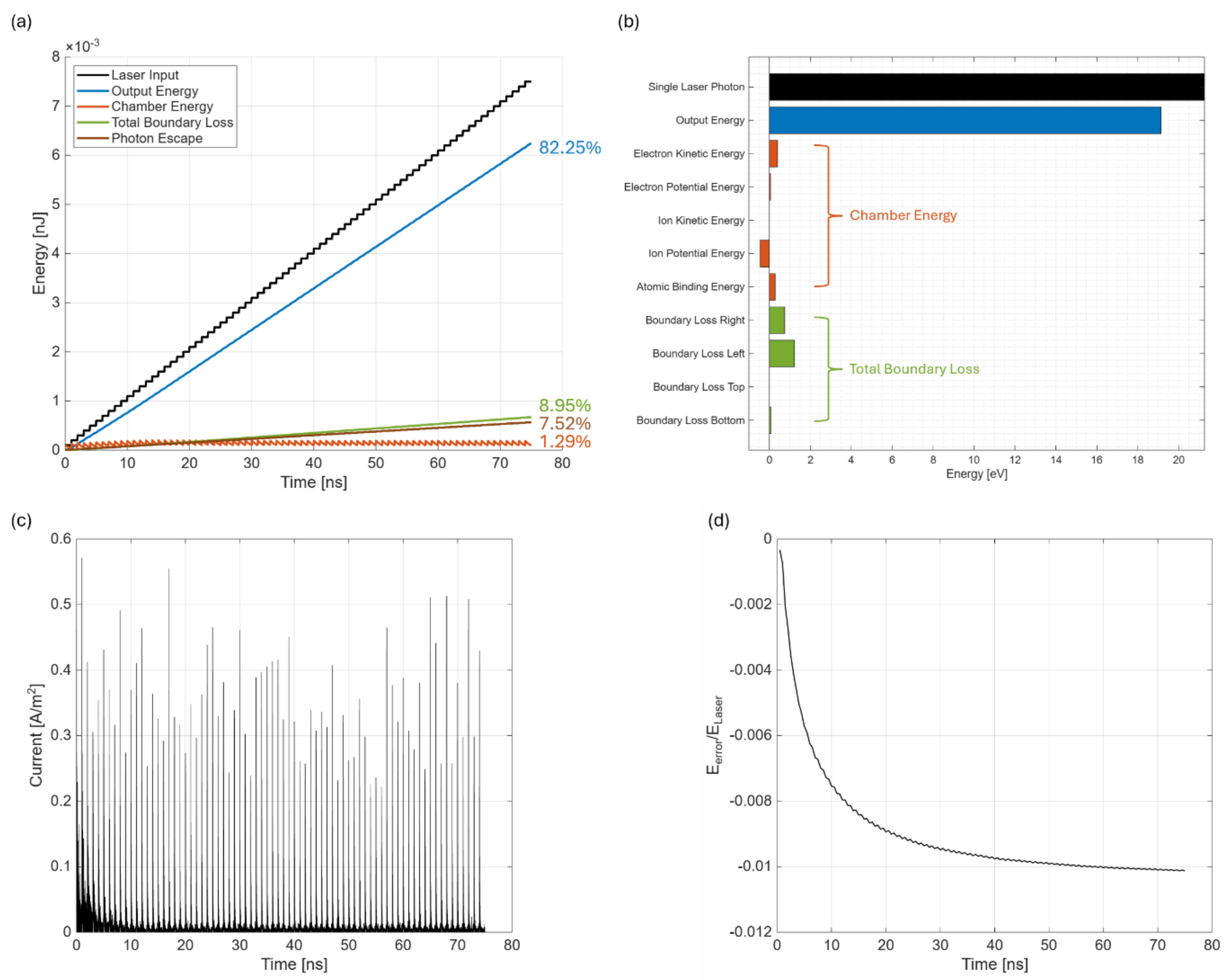

**Figure. 3.** Quantitative energy accounting and steady-state energy partition in the plasma-based laser receiver. (a) The accumulative energy partitions from the initial laser setup to reaching the steady state. (b) The partitions of an absorbed single photon at steady state. (c) The output current density at steady state. (d) The error of energy conservation equation.

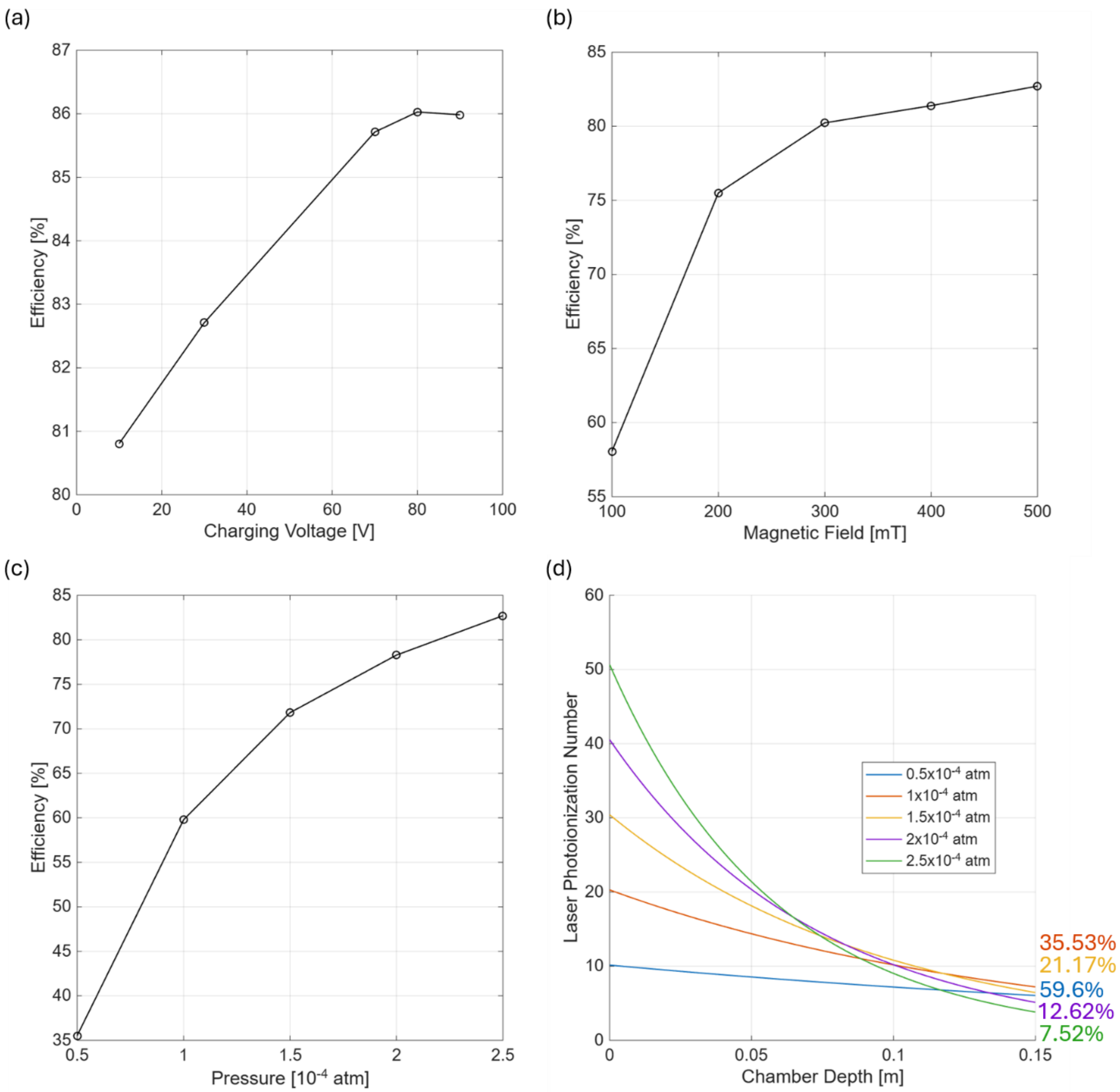

**Figure 4.** Efficiency variation under different simulation conditions. (a) Bias-voltage dependence. (b) Magnetic-field dependence. (c) Pressure dependence. (d) Photon absorption and residual photon number along the chamber depth.

In summary, we numerically demonstrate a plasma-based laser power receiver that converts incident photons into electrical output through gas-phase photoionization, magnetically guided charge separation, and partial internal energy recycling. Self-consistent PIC-MCC simulations predict a receiver-side laser-to-electric efficiency of 82.25% under the representative operating condition, with the energy budget closing within approximately 1%. The remaining energy is accounted for by photon escape, boundary losses, and residual chamber storage. Parameter scans identify photon absorption, electron confinement, and electrostatic extraction as the principal coupled design factors. Although this study is a numerical proof-of-concept requiring further validation of wall interactions, photon recycling, gas retention, and device engineering, it establishes low-pressure plasma as a physically distinct platform for gas-phase optical-to-electrical energy conversion beyond conventional solid-state photovoltaics.

ASSOCIATED CONTENT

Supporting Information

The Supporting Information is available free of charge at ACS Energy Letters.

AUTHOR INFORMATION

**Corresponding Author**

*Michael Keidar, 800 22$^{nd}$ St. NW, Suite 3550, Washington, DC 20052, keidar@gwu.edu

**Author Contributions**

MK and LL contributed to the initial idea. LL contributed to the specific plasma chamber design, PIC-MCC simulation, case sweep and result analysis. ML and LL contributed to the manuscript editing.

**Funding Sources**

This work is supported by the George Washington University (GWU) internal funding.

ACKNOWLEDGMENT

The PIC-MCC simulations were executed in MATLAB and its parallel computing toolbox for the GPU computing, under GWU academic license. Grammarly, Microsoft Copilot, and ChatGPT were used for English-language improvement during manuscript text preparation. All scientific content, analysis, interpretations, and conclusions were developed, verified, and approved by the authors without using AI tools. The authors take full responsibility for the final paper.

ABBREVIATIONS

PIC-MCC particle-in-cell with Monte-Carlo collisions; VUV vacuum ultra-violet; SEE secondary electron emission

## Supporting Information

### Simulation Algorithm and Setups

In the main article, Figure 2(a) summarizes the overall workflow of the self-consistent particle-in-cell Monte Carlo collision (PIC-MCC) simulation used in this work. The simulation resolves the coupled processes of photoionization, charged-particle transport, plasma dynamics, wall interactions, and energy conversion within the xenon chamber. Below are the detailed algorithm and equations of the simulation.

At the beginning of each timestep, the laser pulse is evaluated. When a laser pulse is present, photoionization events are generated according to the local photon absorption probability, producing electron-ion pairs throughout the simulation domain. The amount of photons $N_p$ as the function of chamber depth, a.k.a. laser penetration depth, is computed as

$$N_p(y) = \frac{I_0\tau}{h\nu}\exp(-n_0\sigma y)\,\mathrm{d}x\mathrm{d}z \tag{S1}$$

where $y$ is the chamber depth axis, $I_0$ is the laser pulse power density at 2 MW/m$^2$, $\tau$ is the laser pulse width at 1 ps, $h$ is the Planck's constant, $\nu$ is the photon frequency, $n_0$ is the gas density determined by the chamber pressure, $\sigma$ is the photoionization cross section[1] as shown in Figure 1S, $x$ and $z$ are the chamber width axis.

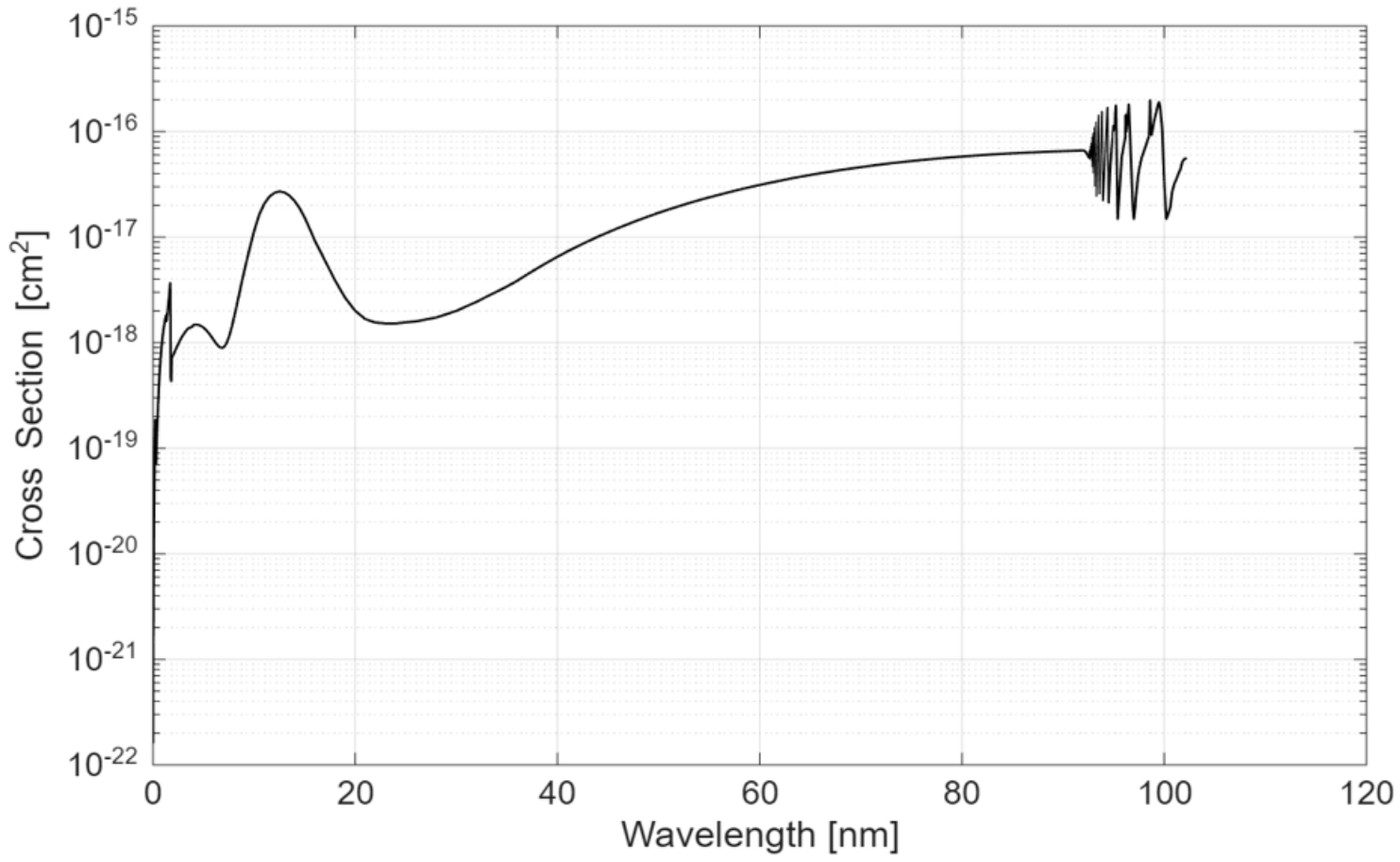

**Figure S1.** The cross section of photoionization on xenon atom

The resulting charge distribution is then used to calculate the net charge density

$$\rho = en_i - en_e \tag{S2}$$

where $e$ is the unit charge, $n_e$ is the electron density, and $n_i$ is the ion density. Next, the electrostatic potential and electric field are obtained by solving Poisson's equation

$$\nabla \cdot \boldsymbol{E} = -\nabla^2\varphi = \frac{\rho}{\varepsilon} \tag{S3}$$

where $\boldsymbol{E}$ is the electric field, $\varphi$ is the electric potential, and $\varepsilon$ is the dielectric constant. The electric field is self-consistently coupled with particle motion and evolves together with plasma.

Charged-particle trajectories are advanced using the Boris particle pusher, which accounts for both electric and magnetic field effects while maintaining excellent numerical stability and energy conservation properties. The charged particle trajectories are advanced using the Boris algorithm, a second-order accurate, time-centered, simplistic integration scheme widely used in particle-in-cell (PIC) simulations due to its excellent long-term conservation properties and unconditional stability in magnetic fields. Particle velocities are staggered by half a timestep with respect to particle positions following the standard leapfrog formulation.

First, a half electric acceleration is computed as

$$\vec{v}^{-} = \vec{v}^{s-\frac{1}{2}} + \frac{q\Delta t}{2m}\vec{E}^{s} \tag{S4}$$

where superscript $s$ represents the time step index rather than the power of the variable, $q$ is the charge of particle, $s-\frac{1}{2}$ represents a half time step, $\Delta t$ is the time step length, and $m$ is the particle mass. Therefore, $\vec{v}^{-}$ is the accelerated particle velocity at the middle of a time step before the magnetic rotation.

Next, magnetic rotation is applied as

$$\vec{\zeta} = \frac{q\Delta t}{2m}\vec{B}^{s} \tag{S5a}$$

$$\vec{\theta} = \frac{2\vec{\zeta}}{1+\left|\vec{\zeta}\right|^{2}} \tag{S5b}$$

$$\vec{v}' = \vec{v}^{-} + \vec{v}^{-} \times \vec{\zeta} \tag{S5c}$$

$$\vec{v}^{+} = \vec{v}^{-} + \vec{v}' \times \vec{\theta} \tag{S5d}$$

where $\vec{B}$ is the magnetic field, $\vec{\zeta}$ is the angle of the rotation in a half time step, $\vec{\theta}$ is the rotational parameter, $\vec{v}'$ is the pre-rotated velocity, and $\vec{v}^{+}$ is the velocity after the magnetic rotation. Next, the second half of electric acceleration that

$$\vec{v}^{s+\frac{1}{2}} = \vec{v}^{+} + \frac{q\Delta t}{2m}\vec{E}^{s} \tag{S6}$$

where $\vec{v}^{s+\frac{1}{2}}$ is the velocity result computed through the Boris pusher based on the Lorentz force. Next, the particle position can be updated as

$$\vec{x}^{s+1} = \vec{x}^{s} + \vec{v}^{s+\frac{1}{2}}\Delta t \tag{S7}$$

After particle transport, Monte Carlo collision (MCC) processes are evaluated, such as electron-impact ionizations, recombination, and other plasma-chemical reactions, etc. as:

$$\frac{\partial n}{\partial t} = \sum_{\xi} k_{\xi} \prod_{\chi} n_{\xi,\chi}^{\delta} \tag{S8}$$

where $n$ is the density of a species, $t$ is time, $k_{\xi}$ is the reaction coefficient of the $\xi^{\mathrm{th}}$ reaction that can either consume or produce the species, and $n_{\xi,\chi}^{\delta}$ is the density of the $\chi^{\mathrm{th}}$ reactant in the $\xi^{\mathrm{th}}$ reaction with the reaction degree of $\delta$. The reaction rate coefficients of pure xenon plasma chemistry are found at LXCat database[2–5].

After the MCC updates on the particle densities, the particles on the wall and outside of the chamber will be removed and the energy losses are recorded as discussed in the main article Eq. (4a) – (4d). First, special treatment is applied to electrons returning from the external circuit. Electrons collected by the output electrode contribute to the electrical output energy, while the return-current electrons are re-injected from the grounded electrode as cold electrons to maintain charge conservation and complete the electrical circuit. These electrons have no drift velocity, in other words, these electrons do not carry energies when injected, meaning that there is no hidden external work. The total number of them is computed from the current output in the previous time step and the distribution of them is uniform in the column of grids next to the left electrode. The entire procedure is repeated for every timestep until the plasma reaches steady-state operation.

**Simulation Mesh Size and Time Step**

The simulation meshing parameters are $\mathrm{d}x = \mathrm{d}y = 0.1$ mm and $\mathrm{d}t = 1$ ps. Considering the low density, such a meshing size is enough to represent the plasma sheath, and the time step is enough to represent the electron drift based on their drift velocity (which will be discussed in the following chapters). Specifically, we have the Debye length $\lambda_D = \sqrt{\frac{\varepsilon k_B T_e}{n_e e^2}} \approx 2.5\ \mathrm{mm} \gg \mathrm{d}x$, and plasma frequency $f_{pe} = \frac{1}{2\pi}\sqrt{\frac{n_e e^2}{\varepsilon m_e}} \approx 100$ MHz leading to the plasma oscillation period $\tau_{pe} = \frac{1}{f_{pe}} \approx 10\ \mathrm{ns} \gg \mathrm{d}t$. Therefore, the meshing size and time step is fine enough to describe the plasma.

**Magnetic Field**

The electrons are magnetized by the magnetic field which is carefully designed to have the exact amplitude to magnetize the electrons. First, we need to make sure that the gyrofrequency of electrons are higher than the collision frequency that $\nu_B = \frac{eB}{2\pi m_e} = 1.4 \times 10^{10}$ Hz $> \nu_c = 2.6 \times 10^7$ Hz. Next, the Larmor radius of electron is $r_e = \frac{m_e v_e}{eB} \sim 0.02$ mm much smaller than the chamber width while the value of ion is $r_i = \frac{m_i v_i}{eB} \sim$ cm. Finally, the magnetic field is also designed with a gradient and curvature to provide a magnetic mirror effect so that the electrons moving to the left will be reflected to the right as shown in Figure S2. Considering the loss cone, we have

$$\sin^2\alpha_{LC} = \frac{B_0}{B_{max}} = \frac{1}{R_m} \tag{S9a}$$

$$\frac{n_{LC}}{n} = 1 - \sqrt{1 - \frac{1}{R_m}} \tag{S9b}$$

where $\alpha_{LC}$ is the angle of the loss cone, $B_0$ is the local magnetic field of an electron, $B_{max}$ is the maximum field magnitude in the simulation domain, $R_m$ is the mirror ratio, $n_{LC}$ is the electron within the loss cone, and $n$ is the total electron number. Recalling that the initial electron velocity is uniformly distributed, while those with initial velocity vector within the loss cone (approximately parallel to the magnetic field) will escape the magnet mirror with the ratio given by Eq. (S9b). However, please note that the electrons escape from the magnetic confinement do not necessarily mean a waste of energy but they may still move towards the right electrodes or will be recaptured by the magnetic field after a collision.

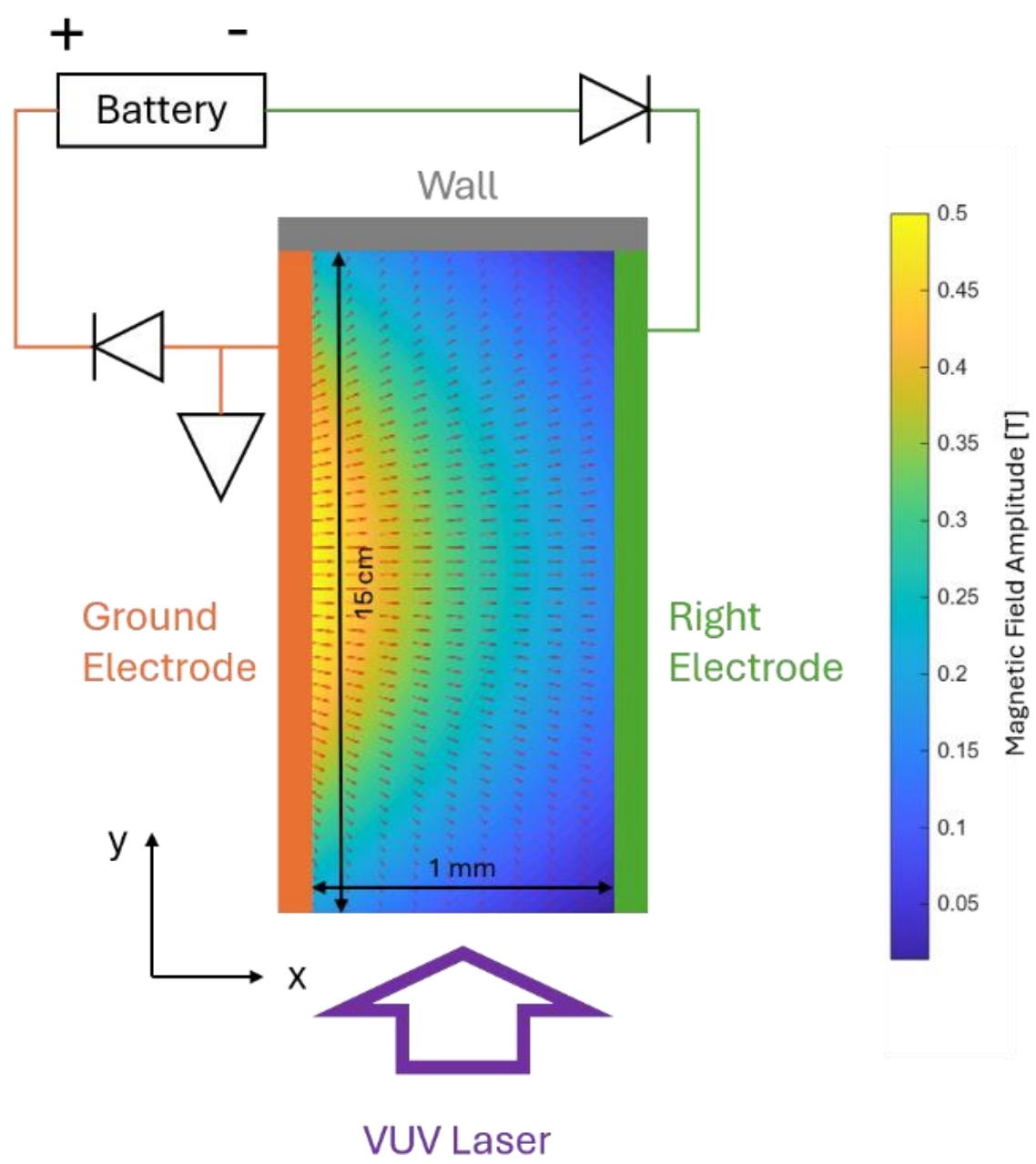

**Figure S2** The magnetic field and chamber design

**Plasma Properties at Steady States**

Figures S3-S14 provide time-resolved snapshots of the plasma properties after the discharge reaches steady-state operation. The four snapshots shown in each figure cover one complete 1000 ps laser period, beginning at 75001 ps immediately after the laser input pulse. Therefore, these plots should be interpreted as intra-period plasma evolution at steady state rather than the initial transient of the whole simulation.

At the beginning of the period, the laser pulse creates a relatively high electron density and ion density. After the pulse, both densities gradually decay as charged particles are extracted by the electrodes, lost at the bottom boundary, and removed through recombination. This periodic generation-and-decay behavior is consistent with the pulsed photoionization source used in the simulation. Because electrons and ions are generated as pairs by photoionization, the electron and ion densities decay in a correlated manner during the laser-off portion of the cycle. As a result, the net charge distribution shown in Figure S5 does not exhibit a significant temporal change over the four snapshots. Small local charge imbalances may appear because electrons and ions have different mobilities and respond differently to the electric and magnetic fields, but the overall plasma remains close to quasi-neutral on the scale resolved in the density plots. Consequently, the space charge is not large enough to reshape the electrostatic potential appreciably.

This weak influence of the net charge is confirmed by Figure S6. The electric potential remains nearly linear from the grounded electrode on the left to the -30 V electrode on the right throughout the laser period. In other words, the potential profile is mainly determined by the externally imposed electrode bias rather than by the transient plasma space charge. The nearly unchanged potential distribution also indicates that the pulsed plasma source modifies the carrier population without strongly perturbing the background electrostatic field used for energy extraction.

Figures S7-S12 present particle-based velocity distributions rather than grid-averaged fields. In these scatter plots, each dot corresponds to one superparticle, and the dot color represents the value of the velocity component indicated by the color bar. The electron velocity components in Figures S7-S9 are much larger than the ion velocity components in Figures S10-S12, as expected from the much smaller electron mass. The electron velocity profiles also show stronger spatial anisotropy because electrons are magnetized under the applied magnetic field and therefore tend to follow helical trajectories along the magnetic-field lines while responding rapidly to the electric field. The magnetic mirror effect provides a physical explanation for the directionality of the electron motion. Electrons moving toward the left encounter an increasing magnetic-field strength, which converts part of their parallel kinetic energy into perpendicular gyromotion and can reflect them before they reach the grounded boundary. This reflection suppresses leftward electron loss and helps maintain a net electron current toward the right output electrode. Although ions are much heavier and their characteristic gyrofrequency is much lower, the ion velocity plots can still show signatures correlated with the magnetic-field geometry. These apparent magnetic effects on ions may arise from direct ion gyromotion when the ion gyro radius becomes comparable to or smaller than the chamber length scale, as well as from indirect coupling through the ambipolar electric field established by the magnetized electron population. Therefore, the ion velocity patterns should be interpreted cautiously: they suggest that ion transport is not completely independent of the magnetic configuration, even though the ion response is much weaker than the electron response.

Figures S13 and S14 show that the electron temperature and gas temperature remain nearly unchanged during the same 1000 ps period. The weak temporal variation of electron temperature indicates that the pulsed photoionization process primarily changes the number of charged particles rather than producing strong intra-period electron heating. The gas temperature is also almost constant because the chamber operates at low pressure, where the collision frequency between charged particles and neutral xenon atoms is low. Under these conditions, only a small fraction of the charged-particle kinetic energy is transferred to the neutral gas within one laser period, so gas heating remains negligible in the steady-state snapshots.

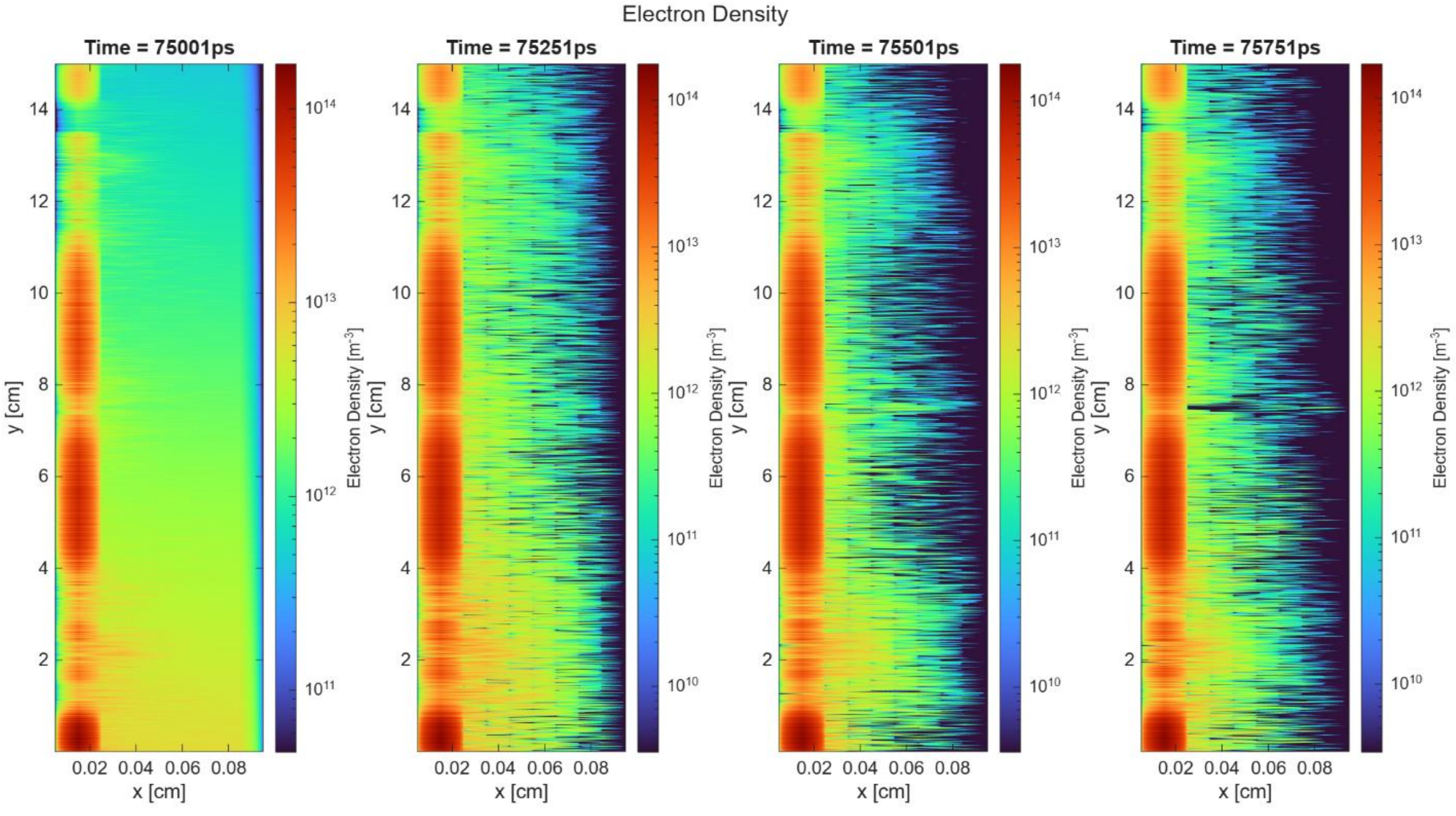

**Figure S3.** Snapshots of the electron density of the pressure at $2.5 \times 10^{-4}$ atm and the maximum magnetic field at 0.5 T

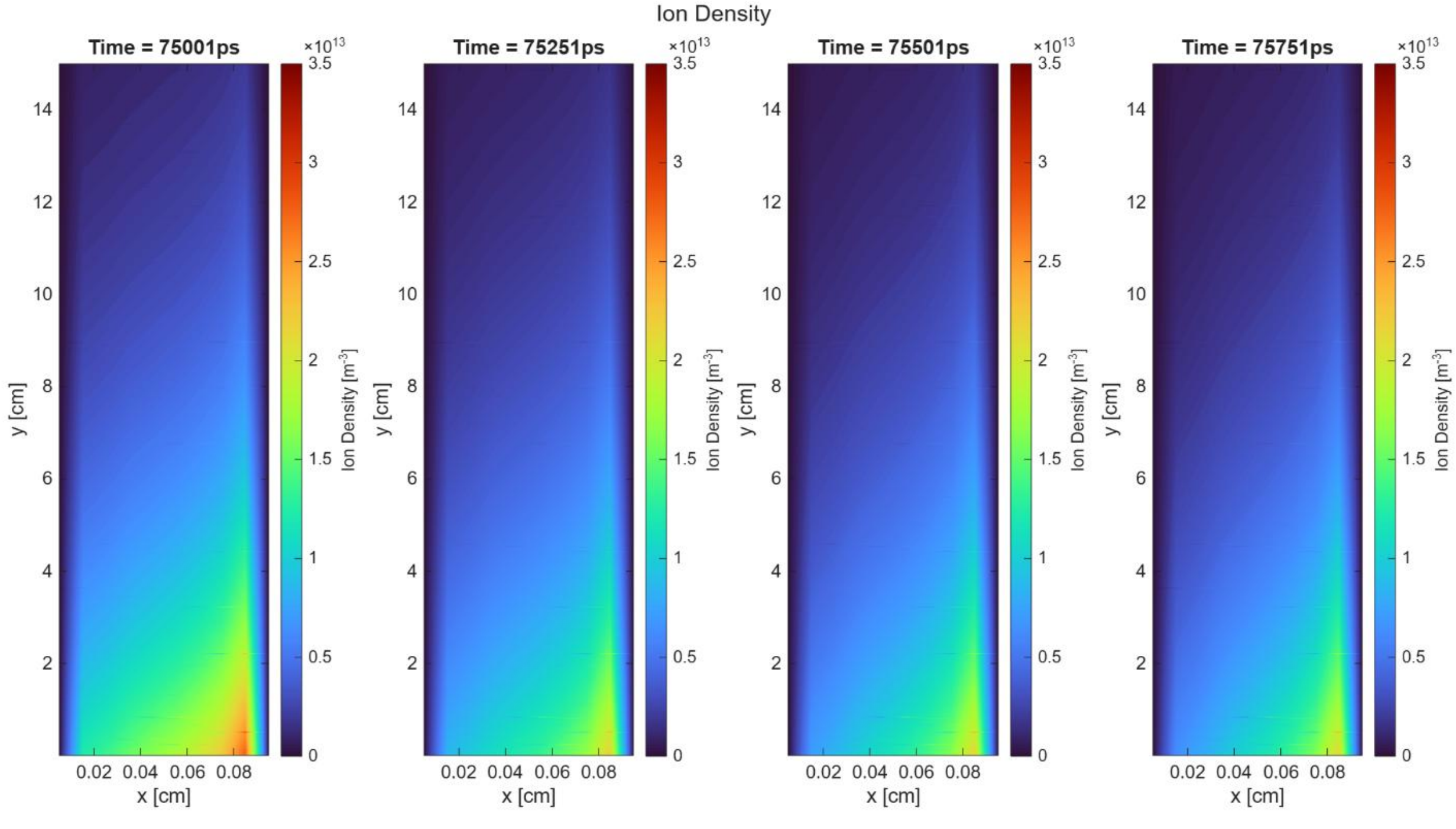

**Figure S4** Snapshots of the ion density of the pressure at $2.5 \times 10^{-4}$ atm and the maximum magnetic field at 0.5 T

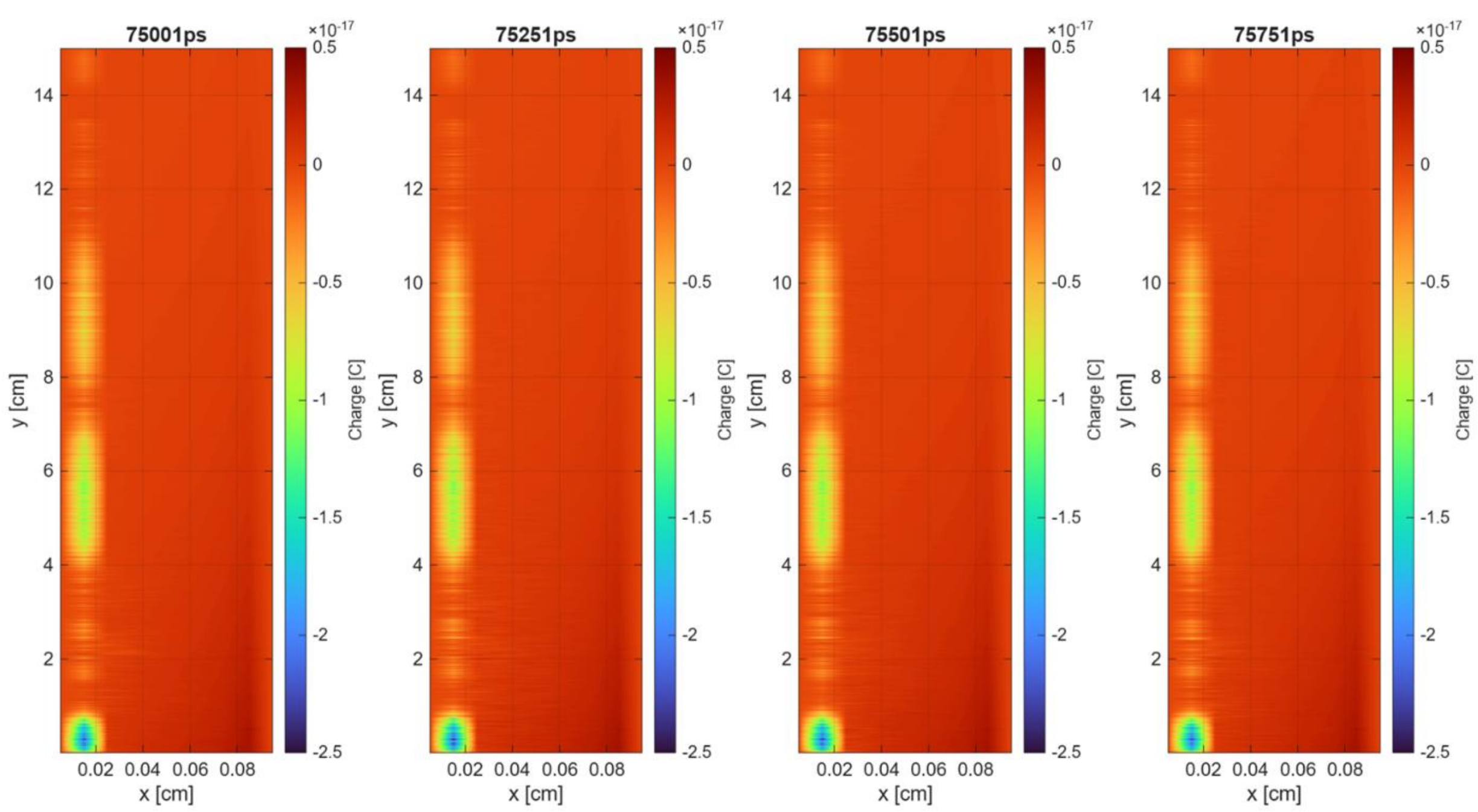

**Figure S5.** Snapshots of the net charge of the pressure at $2.5 \times 10^{-4}$ atm and the maximum magnetic field at 0.5 T

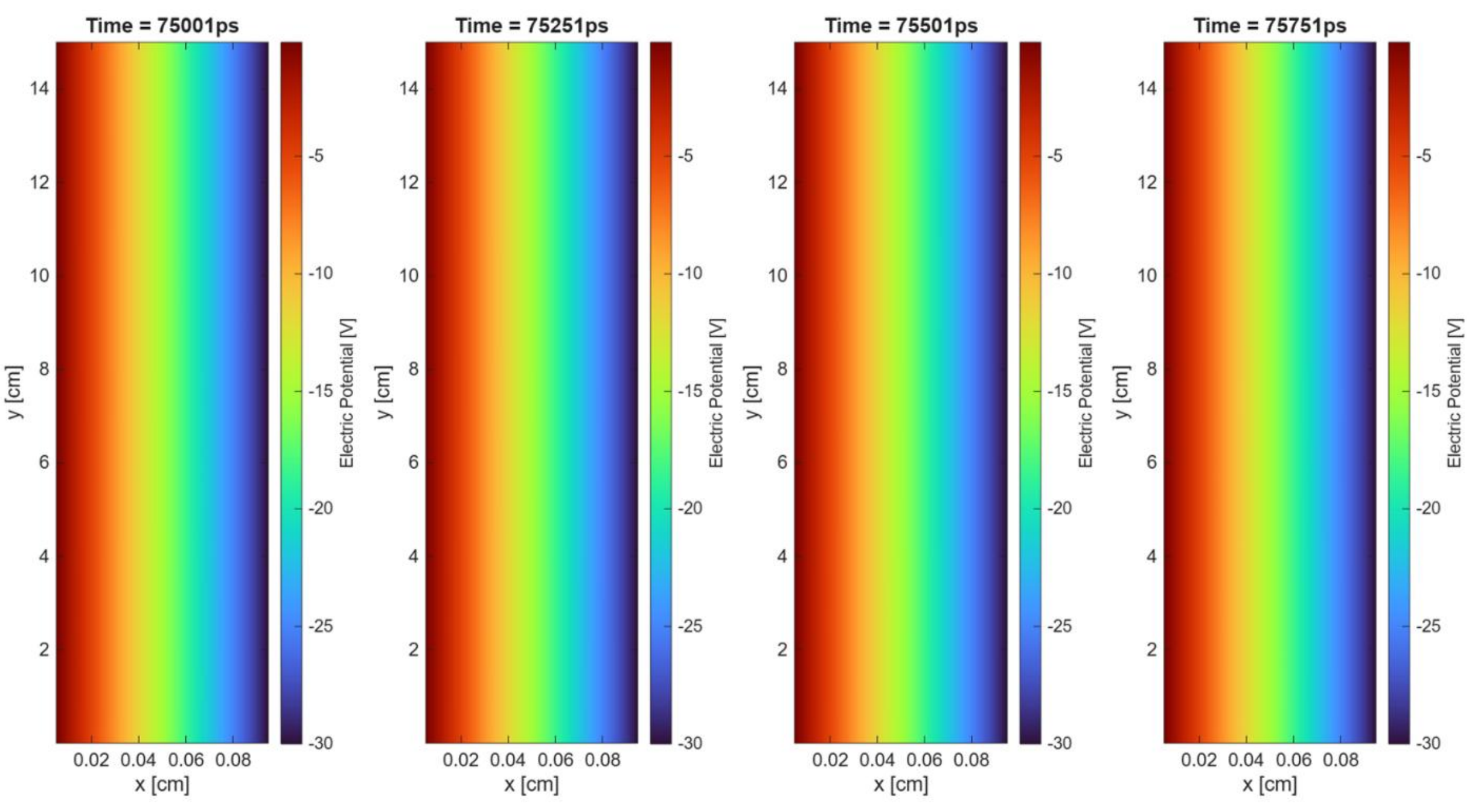

**Figure S6.** Snapshots of the electric potential of the pressure at $2.5 \times 10^{-4}$ atm and the maximum magnetic field at 0.5 T

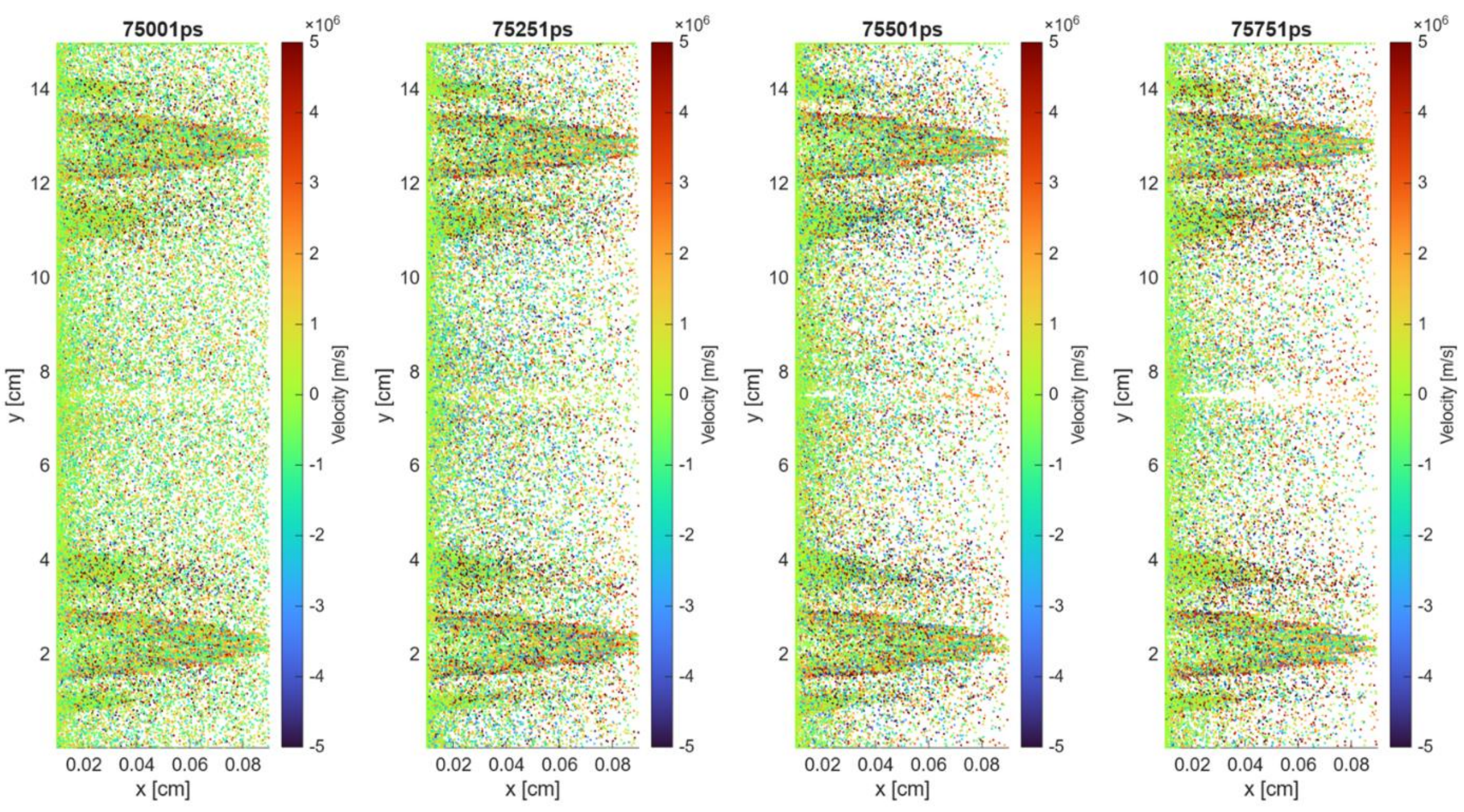

**Figure S7.** Snapshots of the horizontal component of electron velocity of the pressure at $2.5 \times 10^{-4}$ atm and the maximum magnetic field at 0.5 T

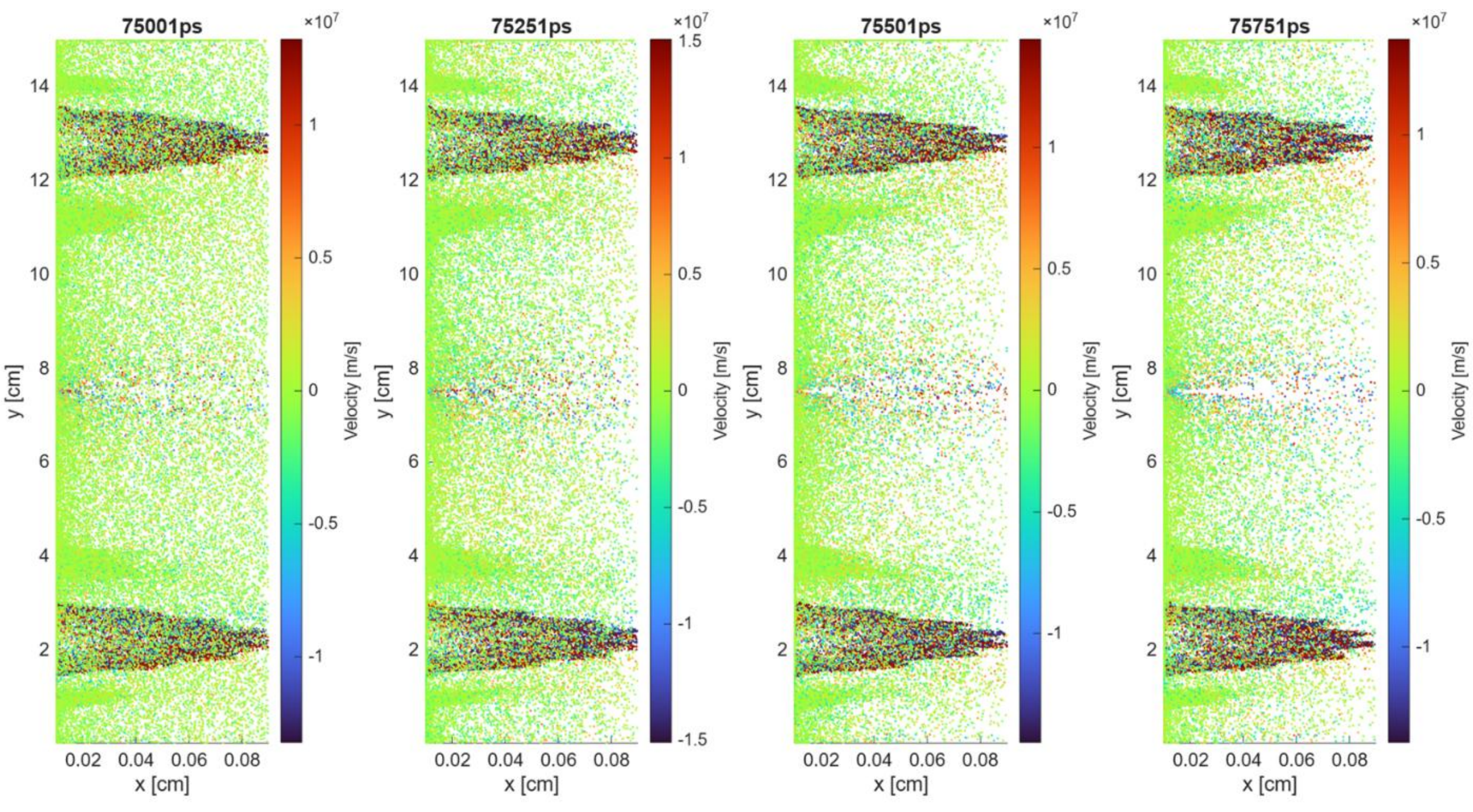

**Figure S8.** Snapshots of the vertical component of electron velocity of the pressure at $2.5 \times 10^{-4}$ atm and the maximum magnetic field at 0.5 T

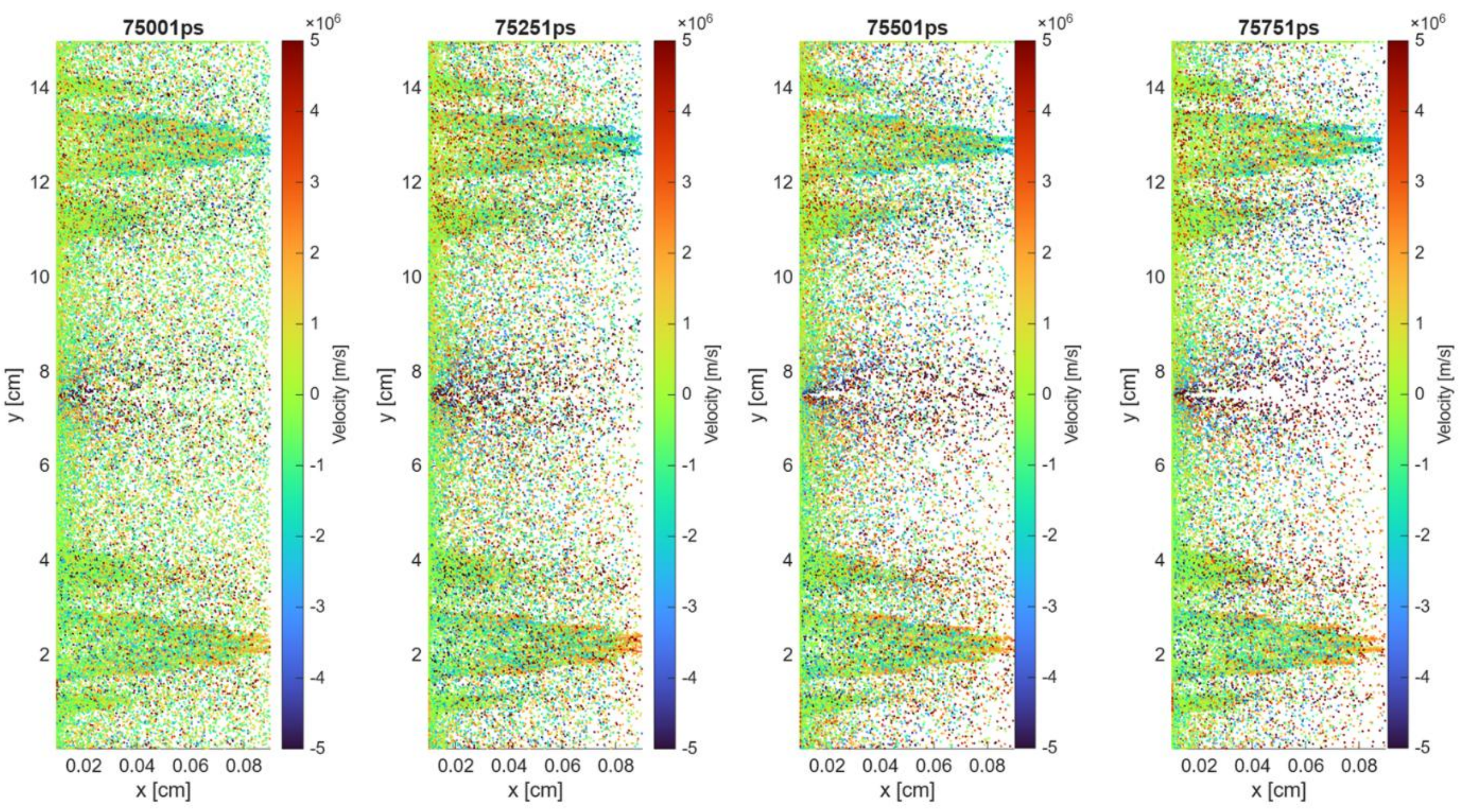

**Figure S9.** Snapshots of the outbound component of electron velocity of the pressure at $2.5 \times 10^{-4}$ atm and the maximum magnetic field at 0.5 T

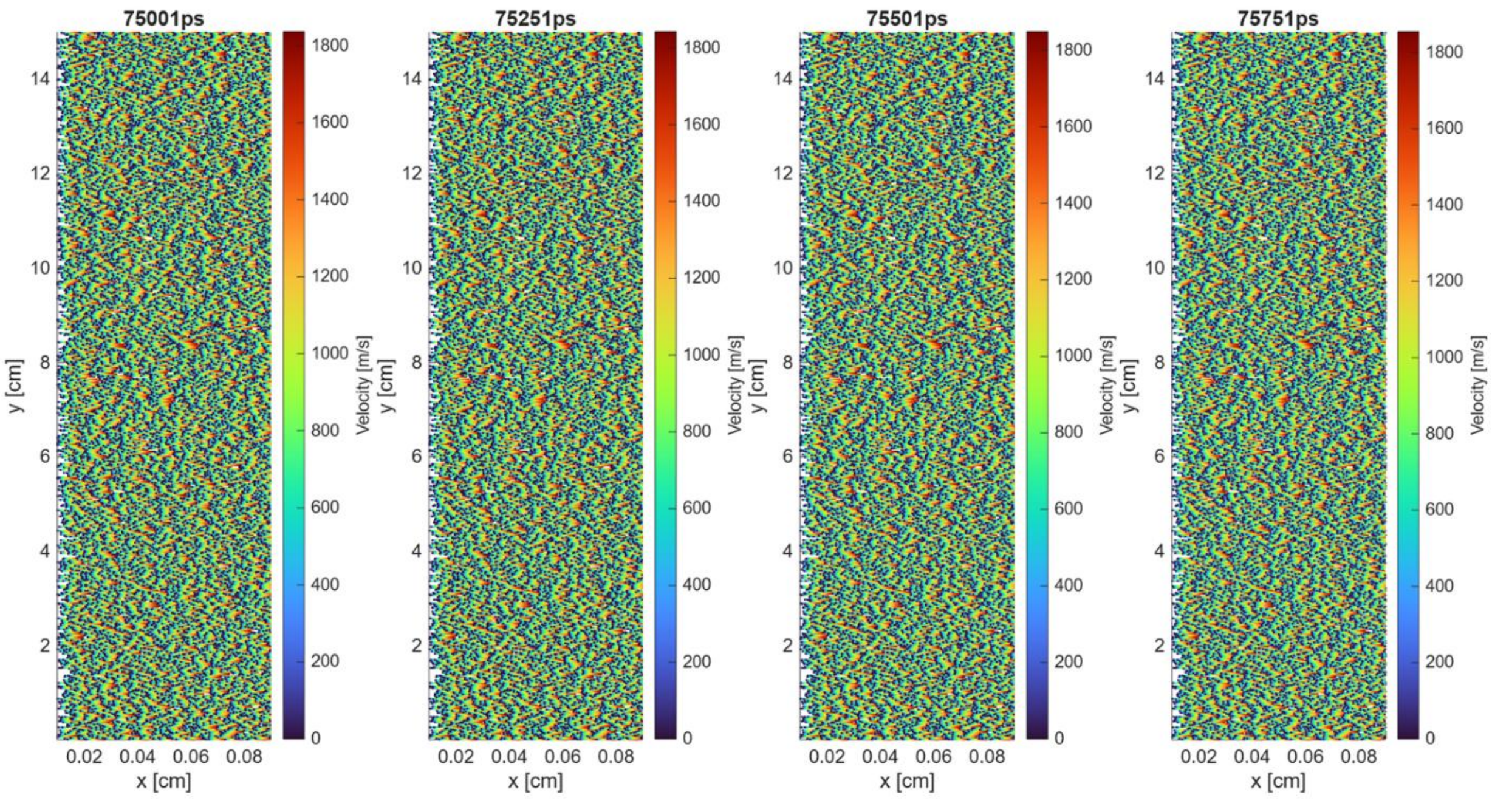

**Figure S10.** Snapshots of the horizontal component of ion velocity of the pressure at $2.5 \times 10^{-4}$ atm and the maximum magnetic field at 0.5 T

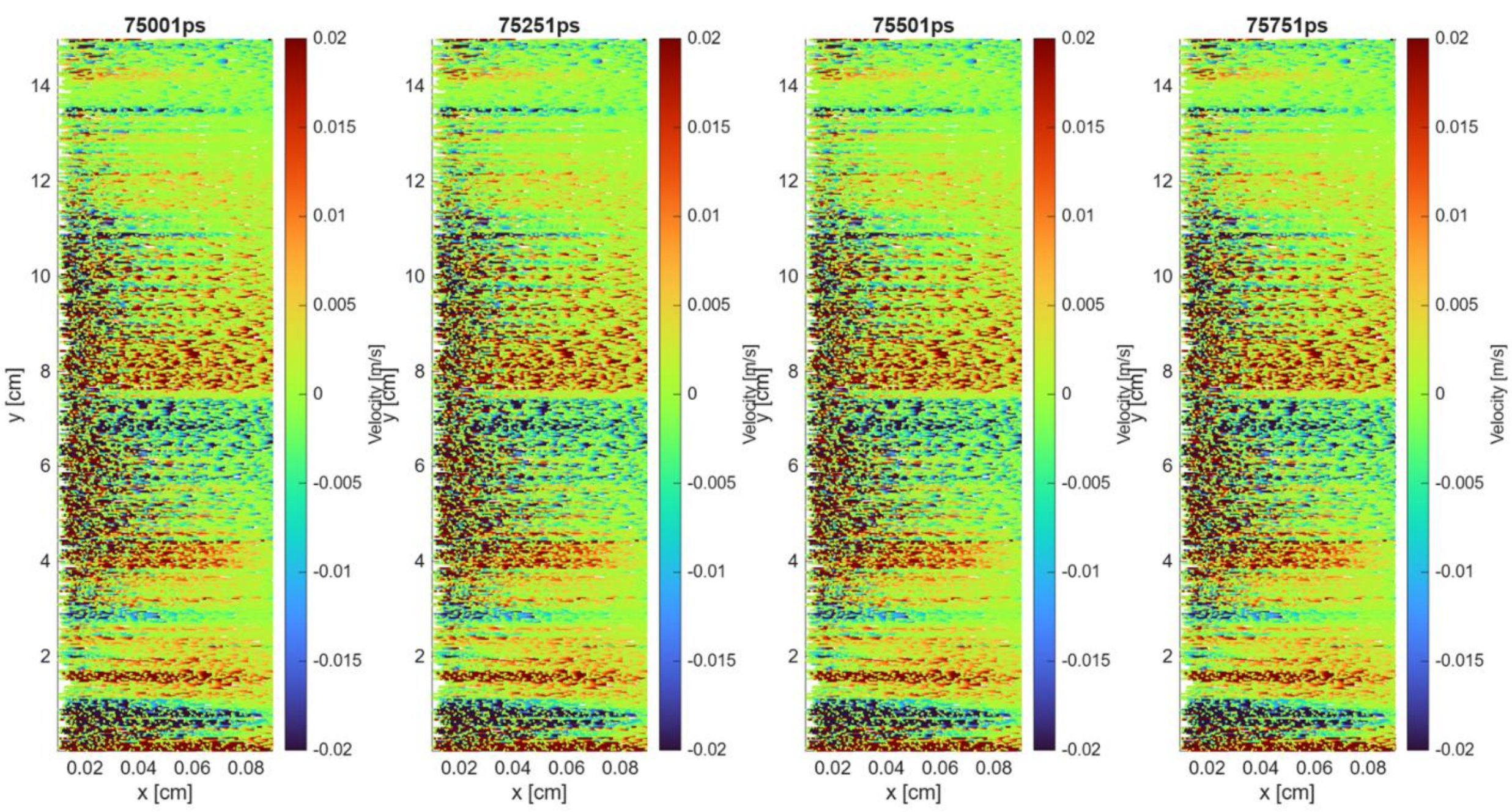

**Figure S11.** Snapshots of the vertical component of ion velocity of the pressure at $2.5 \times 10^{-4}$ atm and the maximum magnetic field at 0.5 T

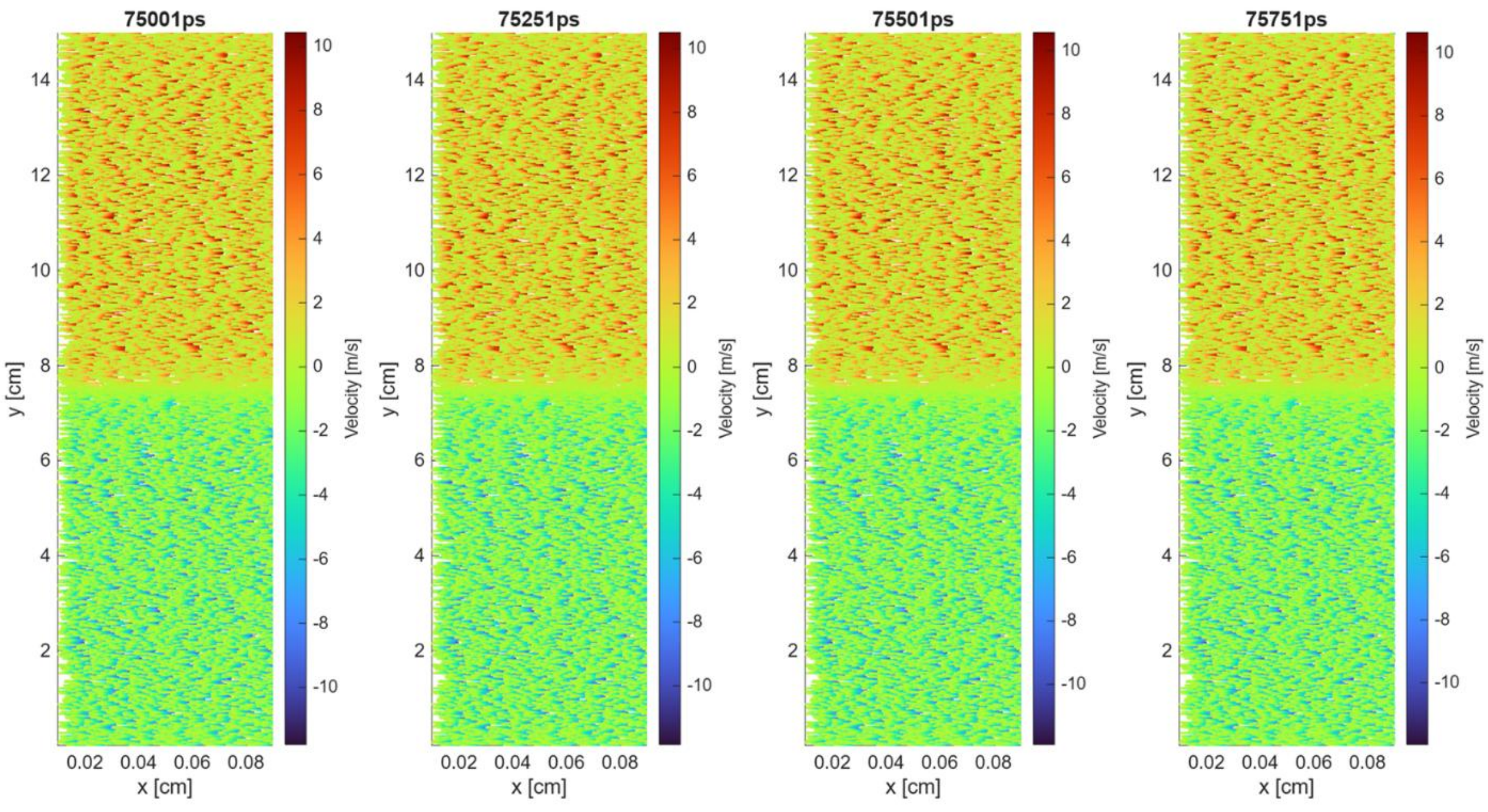

**Figure S12.** Snapshots of the outbound component of ion velocity of the pressure at $2.5 \times 10^{-4}$ atm and the maximum magnetic field at 0.5 T

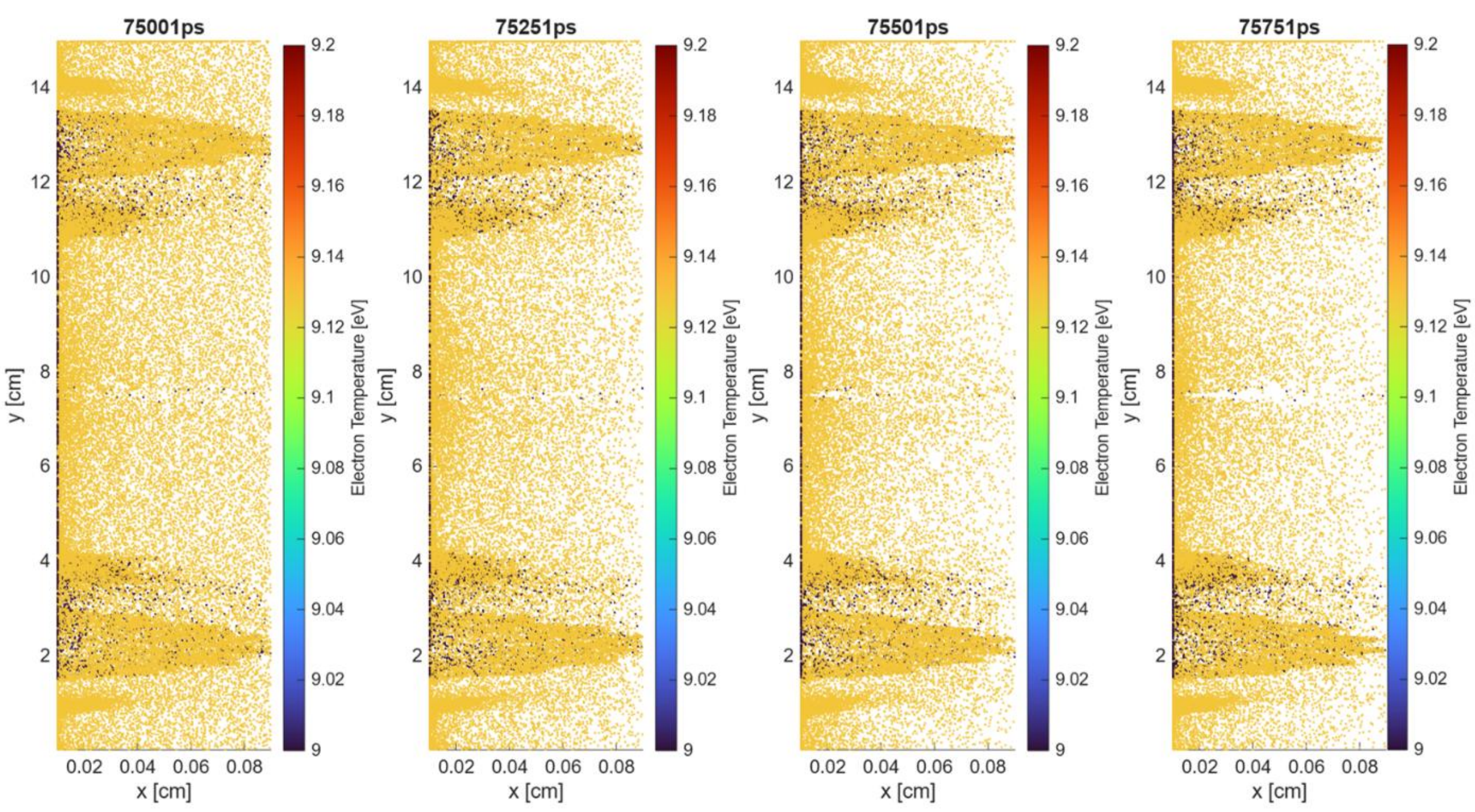

**Figure S13.** Snapshots of the electron temperature of the pressure at $2.5 \times 10^{-4}$ atm and the maximum magnetic field at 0.5 T

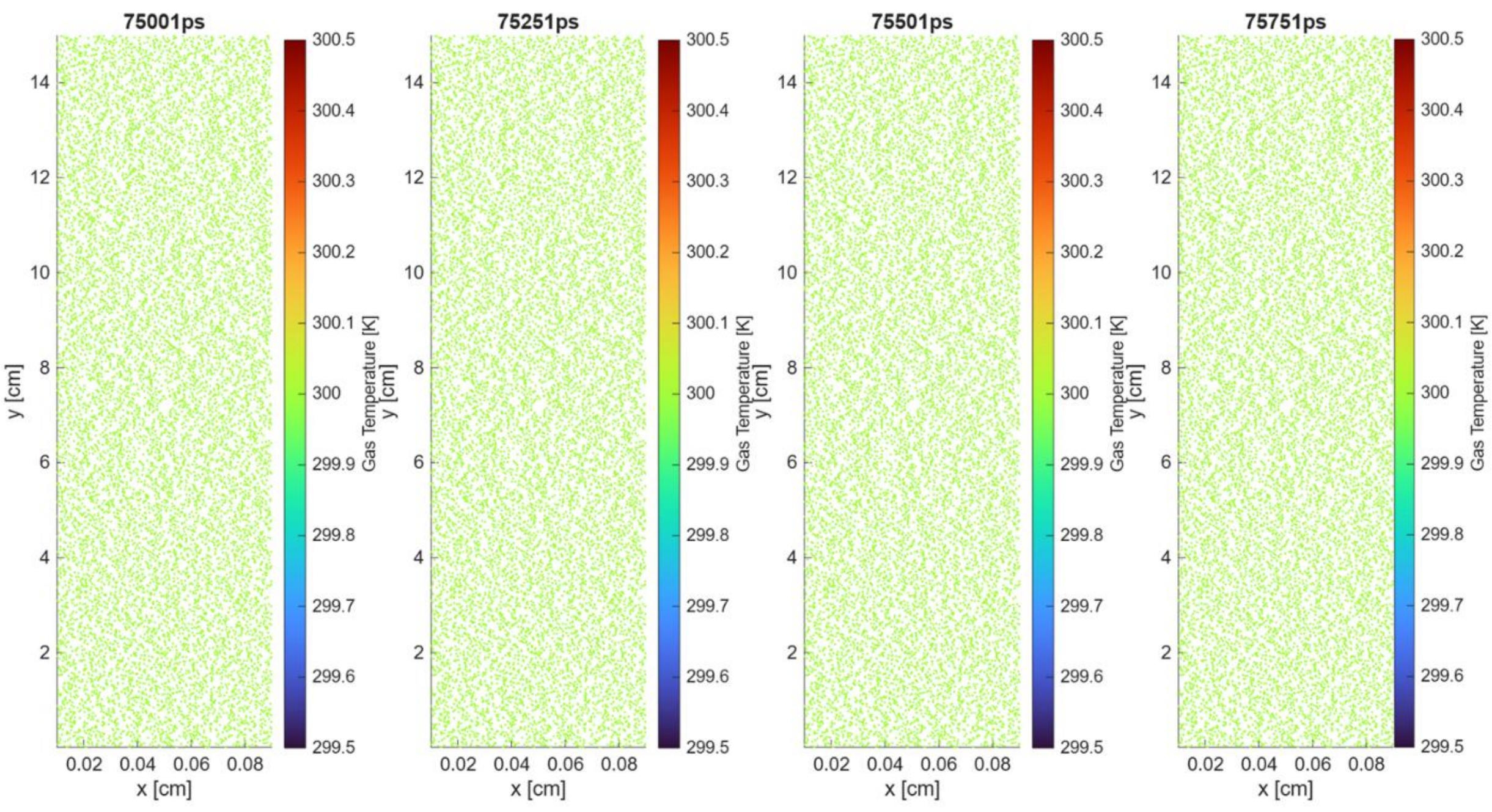

**Figure S14.** Snapshots of the gas temperature of the pressure at $2.5 \times 10^{-4}$ atm and the maximum magnetic field at 0.5 T

## Secondary Electron Emission (SEE) Model

The Vaughan-type empirical SEY model is given in Eq. (S10a) – Eq. (S10c). In this model, $E_e$ is the impact electron kinetic energy, $E_{SEEth}$ is the threshold energy for SEE emission and is taken as 10 eV, and $E_{max}$ is the empirical fitting energy at which the SEE yield reaches its maximum, taken here as 400 eV. In Eq. (S10c), $f_{SEE}$ is the energy spectrum of SEE that can be found at the reference15. The resulting $\delta_{SEE}$ represents the number of secondary electrons generated near the wall by an incident electron; physically, these emitted electrons recover part of the impact electron energy by redistributing it into newly generated near-wall electrons whose total kinetic energy corresponds to the recoverable portion of the incident electron kinetic energy. $\delta_{SEE} > 1$ means the number of SEE generated by an electron impact while $\delta_{SEE} < 1$ means the probability to generate an SEE.

$$\chi = \max\left\{\frac{E_e - E_{SEEth}}{E_{max} - E_{SEEth}}, 0\right\} \quad \text{(S10a)}$$

$$\delta_{SEE} = \frac{1.7\chi}{0.7 + \chi^{1.7}} \quad \text{(S10b)}$$

$$E_{SEE} = \int_0^{E_e} E f_{SEE}(E)\mathrm{d}E \quad \text{(S10c)}$$

## Bottom Boundary: The Laser Input Window

As mentioned in the main article, the laser input window is an opening rather than a material transparent to the VUV laser. The opening size is 1 mm which is less than the mean free path of the gas at $2.5 \times 10^{-4}$ atm pressure. Also, a slow gas supply to the chamber can be considered either during the charging or between charging periods. The gas loss rate can be computed using Knudsen effusion equation that

$$\dot{m} = AP\sqrt{\frac{m}{2\pi k_B T_g}} \quad \text{(S11)}$$

where $A$ is the area of the opening window, $P$ is the pressure, $m$ is the mass of the particle, $k_B$ is the Boltzmann constant, and $T_g$ is the gas temperature. The result is shown in Figure S15. The results show that even for the gas leak rate at about 2.5 grams per day with $1 \times 10^{-3}$ atm pressure, higher than the cases discussed in this work, considering a 2 kg gas tank commonly equipped by electric propulsion and satellites [6,7], the time to deplete the supply will be years. This proves that the bottom window, with its size less than the mean free path of the gas, has an infinitesimal gas loss rate. Please note that the energy loss rate due to the escape of photons, electrons, and ions are already considered in the main article. The above estimation of Knudsen effusion equation is used to verify the engineering feasibility of the bottom window design.

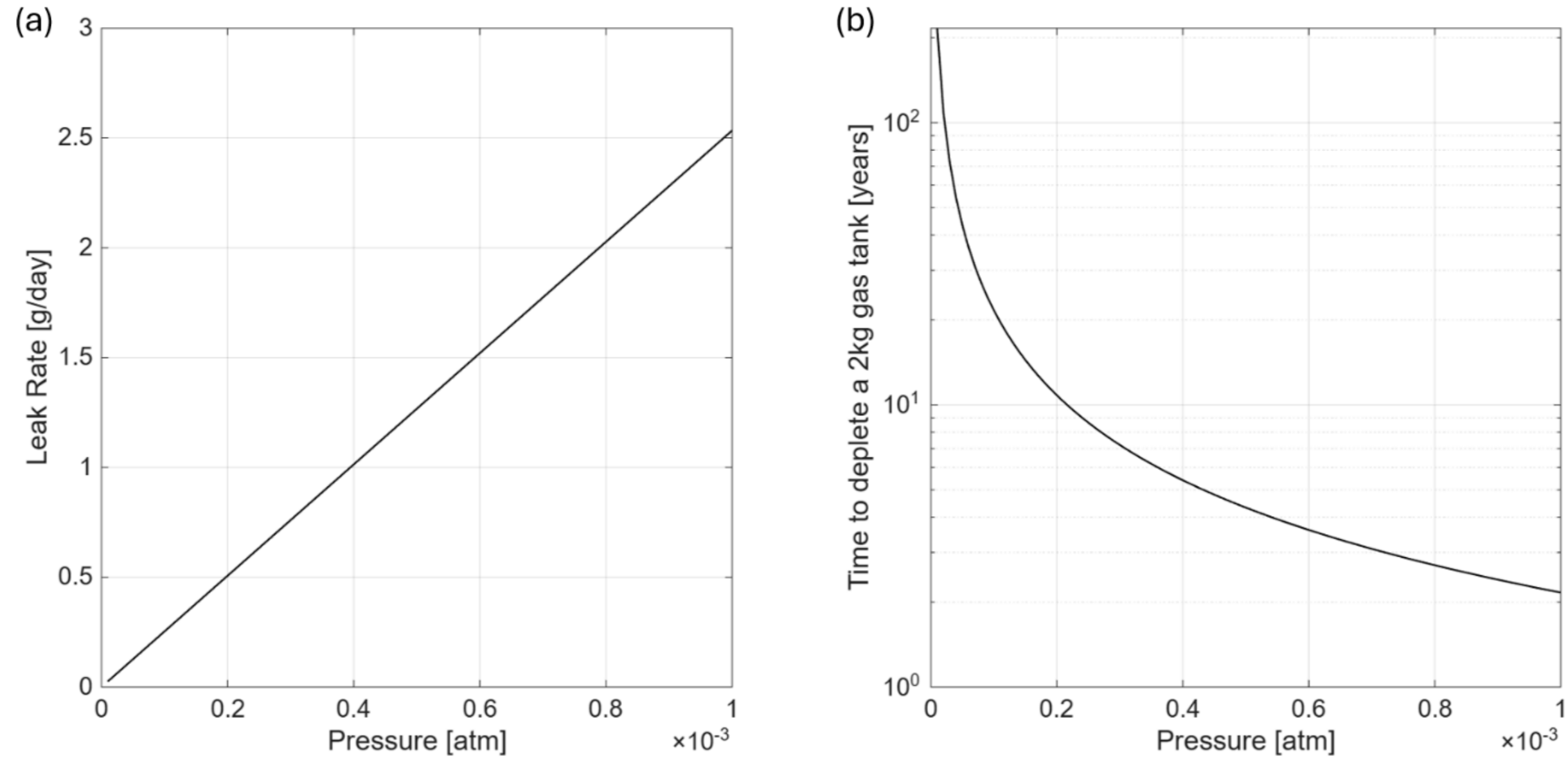

**Figure S15.** The leak of gas through the chamber bottom window where the size is smaller than the mean free path. (a) The leak rate. (b) The time to deplete a 2 kg gas tank.